\documentclass[aps,
prd,
preprint,
eqsecnum,
amsmath,
amssymb
]{revtex4}

\usepackage{hyperref}
\usepackage{bm}
\usepackage{color}
\usepackage{xcolor}

\begin{document}

\title{Constrained field theories on Kerr backgrounds} 

\author{Karan Fernandes} \email[email: ]{karan12t@bose.res.in}
\affiliation{S N Bose National Centre for Basic Sciences, Block JD, Sector III, Salt Lake, Kolkata 700106, India}

\author{Amitabha Lahiri} \email[email: ]{amitabha@boson.bose.res.in}
\affiliation{S N Bose National Centre for Basic Sciences, Block JD, Sector III, Salt Lake, Kolkata 700106, India}

\begin{abstract}

We analyze the constraints of gauge theories on Kerr and Kerr-de Sitter spacetimes, 
which contain one or more horizons. We find that the constraints are modified on such 
backgrounds through the presence of additional surface terms at the horizons. 
As a concrete example, we consider the Maxwell field and find that the Gauss law 
constraint involves surface corrections at the horizons. These surface contributions 
correspond to induced surface charges and currents on the horizons, which agree with 
those found within the membrane paradigm. The modification of the Gauss law constraint also 
influences the gauge fixing and Dirac brackets of the theory.

\end{abstract}

\pacs{}

\maketitle

\section{Introduction}
The horizons of black holes are a profound consequence of the General Theory of Relativity.
Black holes present to the universe 
a closed surface of finite size, completely characterized by macroscopic parameters such as mass, 
charge and spin~\cite{Chandrasekhar:1985kt}.  Information about the internal structure 
of a black hole is unobservable from the outside due to the presence of the horizon, at least classically. 
The seminal discovery by Hawking~\cite{Hawking:1974sw} that a black hole radiates like a black body 
with a finite temperature, following Bekenstein's suggestion that a black hole possesses an entropy 
proportional to the surface area of its horizon~\cite{Bekenstein:1973ur}, implies the possibility that a 
black hole has associated with it a very large number of microscopic 
states. It is natural to think that these states are in some way related to the degrees of 
freedom of the horizon. This view has been strengthened in approaches that treat fields on black hole 
backgrounds as those of manifolds with boundaries. For gravity, this approach leads to a quantum 
description in which an infinite set of observables are localized on the 
boundary~\cite{Balachandran:1994up, Balachandran:1995qa, Carlip:1998wz, Carlip:1999cy}. 

There has been a resurgence of interest in 
studying the behaviour of quantum fields near black hole horizons, motivated by various 
paradoxes and puzzles related to the information problem~\cite{Almheiri:2012rt, Braunstein:2009my}. 
Based on the asymptotic symmetries of fields on the null boundaries of conformally compactified flat
spacetimes~\cite{Balachandran:2013wsa, Campiglia:2014yka, Strominger:2013lka, He:2014cra, 
	Strominger:2013jfa, He:2014laa}, there have been recent proposals for the existence of soft 
black hole hairs~\cite{Hawking:2016msc,Afshar:2016wfy,Mirbabayi:2016axw,Hawking:2016sgy,
Tamburini:2017dig}. The significance of the horizon is highlighted in the 
membrane paradigm, where one replaces the black hole by a 
membrane with certain classical properties at the stretched horizon, i.e. a small distance 
outside the event horizon (an excellent overview is provided by the collection of articles 
in~\cite{Thorne:1986iy}). This is a sensible description from the perspective of an external 
stationary observer, who finds that particles cannot classically leave the interior of the 
black hole or reach the horizon from the outside in finite time. Thus the classical or semi-classical 
dynamics of fields on black hole backgrounds may be studied by considering the bulk and the horizon, 
and completely ignoring what happens in the interior of horizon.

Boundary conditions on the fields play a crucial role in all these investigations. 
In most of these papers, though not all of them, the fields (or their derivatives) 
are set to vanish on the horizon. For many field theories, this is a convenient 
way of ensuring that invariants constructed out of the stress energy tensor remain 
finite at the horizon. For the Kerr black hole spacetime, boundary conditions on the 
components of electric and magnetic fields relate the charge and surface currents 
at the horizon~\cite{Znajek:1978,Damour:1978cg, Price:1986yy}. 
These conditions allow for the extraction of electromagnetic energy from Kerr black 
holes through a magnetic Penrose process~\cite{Blandford:1977ds}.
The boundary conditions for gauge fields are 
special in that we can ensure the finiteness of gauge invariant observables 
without necessarily imposing the finiteness on the components of gauge fields. 
In addition, assuming any particular values for gauge fields is not particularly 
meaningful, as they are defined always up to gauge transformations.

Gauge theories are characterized by the presence of redundant degrees of 
freedom, which leads to the presence of constraints. The formalism for studying
the dynamics of constrained systems was discovered by Dirac~\cite{Dirac:1950pj}
and independently by Bergmann et al.~\cite{Bergmann:1949zz,Anderson:1951ta}, and has 
been applied to numerous theories of interest over the 
years~\cite{Dirac-lect-1964,Hanson:1976cn,Henneaux:1992ig}. 
While the formalism for constrained field theories set up by Dirac generalizes to 
curved backgrounds~\cite{Dirac:1951zz},  the more general formulation in terms of shift 
and lapse variables was introduced by Arnowitt, Deser and Misner~\cite{Arnowitt:1962hi}. 
In particular, this formulation has been used to understand the initial value problem 
of fields theories~\cite{Isenberg:1977ja}, the behaviour of the fields near the horizons 
of stationary black hole spacetimes~\cite{MacDonald:1982zz}, and its quantization~\cite{Sorkin:1979ja}. 
Until recently, a noticeable absence in the literature involved the formulation of constrained theories 
on curved backgrounds with horizons. The modification of constraints due to spatial boundaries 
on flat backgrounds were investigated in~\cite{SheikhJabbari:1999xd,Zabzine:2000ds}, while 
in~\cite{Balachandran:1993tm,Balachandran:1992qg} the quantization of the Chern-Simons 
theory on a disk and the role of boundaries on the vacuum structure of the theory has been 
covered in detail. It is the boundary conditions on gauge fields at the horizon that concerns us
in this paper. The point is that the value of a gauge field at a boundary can be changed by a 
gauge transformation. The only way to fix the boundary value of a gauge field is to restrict 
to gauge transformations which vanish at the boundary. However, there is no sensible reason to 
do that when the said boundary is not a physical singularity, so it is sufficient to keep the 
gauge transformations regular at the horizon. We will find that this seemingly innocuous 
condition leads to a modification of the system of constraints when a horizon is present. 

The formulation of gauge theories on spherically symmetric backgrounds 
with horizons was considered in~\cite{Fernandes:2016imn}, where it was found that the 
constraints received contributions from terms localized at the horizon. In particular, this is 
true for the Gauss law constraint in electrodynamics, which now has an additional 
contribution from the horizon. This resulted in a vanishing charge for an observer 
situated at the horizon of a Reissner-Nordstr\"om
black hole, while not affecting the usual charge observed by the oberver at infinity. 
In the present work we investigate the classical constraints of electrodynamics in Kerr spacetimes. 

The organization of our paper is as follows. In Sec.~\ref{geom}, we set up our 
notations and conventions for the analysis of constraints on the Kerr background.  
In Sec.~\ref{Max}, we consider Maxwell's theory and explicitly derive 
the surface contributions to the constraint on the horizon. Gauge fixing is 
considered in both the radiation gauge and axial gauge.
Finally in Sec.~\ref{Con}, we discuss the physical consequences of our results.
This involves a modification of the usual solution for the electromagnetic 
scalar potential known in the absence of boundaries.

\section{General Algorithm} \label{geom}
\subsection{Kerr backgrounds}
Here we consider the description of 
hypersurfaces for the Kerr background which will be needed in our treatment of constrained field theories. 
The spacetime, which may possess one or more horizons (as in the Kerr-de Sitter case) admits two Killing vector fields:
a stationary $\xi^a$ and an axial $\omega^a$\,, whose normalization we take to be 
\begin{align}
\xi_a \xi^a &= -\lambda^2 \, ,\notag\\
\omega^a \omega_a &= f^2 \, .
\label{gen.Kill}
\end{align}
The orbits of $\omega^a$ are taken to be closed, i.e. $\omega^a$ is periodic. The Killing vector fields mutually commute with each other, 
\begin{equation}
\left[\xi,\omega\right]^a = \xi^b\nabla_b\omega^a - \omega^b \nabla_b \xi^a = 0 \, .
\label{gen.comm}
\end{equation}
Kerr backgrounds admit spatial hypersurfaces which are tangent to $\omega^a$ and orthogonal 
to the vector (but not Killing) field
\begin{equation}
\chi^a = \xi^a + \alpha \omega^a \, ,
\label{gen.chi}
\end{equation}
where $\alpha$ is defined through the contraction of the Killing vectors
\begin{equation}
\alpha = -\frac{1}{f^2}\xi_a \omega^a  \, .
\label{gen.alpha}
\end{equation}
We note that $\alpha$ in general is not a constant. It now follows that this vector is timelike in the region where $\lambda^2 + \alpha^2 f^2$ is positive, since
\begin{equation}
\chi_a \chi^a = -\beta^2 = - (\lambda^2 + \alpha^2 f^2)
\label{chi.norm}
\end{equation}
Despite being a combination of Killing vectors, $\chi^a$ itself is not Killing since
\begin{equation}
\pounds_{\chi} g_{cd} = 2 \omega_{(c} \nabla_{d)} \alpha \, .
\label{chi.nLie}
\end{equation}
$\chi^a$ only coincides with the Killing vector on those surfaces where $\beta^2=0$ and $\alpha$ is a constant, i.e. on the horizons of the spacetime~\cite{Bhattacharya:2011dq}. Hence $\chi^a$ is timelike in the region outside the event horizon of asymptotically flat backgrounds, or in the general case of black hole de Sitter backgrounds, between the event horizon and the cosmological horizon.

It is straightforward to verify that $\chi^a$ satisfies the Frobenius condition
\begin{equation}
\chi_{[a} \nabla_{b} \chi_{c]} = 0 \, .
\label{chi.Frob}
\end{equation}
Thus $\chi^a$ is a timelike vector which is orthogonal to some spatial hypersurface 
$\Sigma$. Since $\chi^a\omega_a = 0$, these hypersurfaces are also tangent to the Killing vector $\omega^a$. 
The projection operator on $\Sigma$ is given by
\begin{align}
h^{a}_{b} = \delta^{a}_{b} + {\beta}^{-2} \chi^{a} \chi_{b}\, .
\label{gen.met}
\end{align}
We assume that the spacetime is `Kerr-like', i.e., an orthonormal basis on it is $\{\beta^{-1} \chi^a\,, f^{-1} \omega^a\,,
\mu^a\,,\nu^a\}\,,$ where the unit vectors $\{\mu^a\,,\nu^a\}$ are
orthogonal to both  $\xi^a$ and $\omega^a$ (and it follows, to $\chi^a$) and
span an integral submanifold. Both the Kerr and Kerr-de Sitter spacetimes fall in this category.
Using this basis, we can express  the spatial projector given in Eq.~(\ref{gen.met}) as
\begin{equation}
h^a_b = f^{-2}\omega^a\omega_b + \mu^a\mu_b+ \nu^a\nu_b\,.
\label{gf2.proj2}
\end{equation}
The Killing horizons $\cal{H}$ are closed, axially symmetric surfaces, which are submanifolds of $\Sigma$. The induced metric on $\cal{H}$ is given by
\begin{equation}
\sigma_{ab} = h_{ab} - n_a n_b \, ,
\label{hor.proj}
\end{equation}
where $n^a$ is the outward (inward) pointing unit spatial normal to the inner (outer) horizon of the background, satisfying $n_a n^a = 1$. Since the horizon is axially symmetric and $\omega^a$ is tangent to the hypersurface $\Sigma$, it also follows that $n_a \omega^a = 0$.

Using $h^a_b$ we can now project any spacetime tensor onto the spacelike hypersurface $\Sigma$. Denoting the covariant derivative on the hypersurface by $\mathcal{D}_a = h_a^b \nabla_b$, we have the following projection
\begin{equation}
\mathcal{D}_a t^{c...e}_{b...d} = h^{a'}_a h^{b'}_{b} h^{c}_{c'} \cdots h^{e}_{e'} h^{d'}_{d} \nabla_{a'} T^{c'...e'}_{b'...d'}\,,
\label{gen.proj2}
\end{equation}
where $T^{c...e}_{b...d}$ is a spacetime tensor and $t^{c...e}_{b...d}$ denotes its projection on the spacelike hypersurface.

The time coordinate is measured along $\chi_a$ and is constant on the hypersurface $\Sigma$. In what follows we will consider the time evolution vector to be along $\xi^a$. With this choice, $\alpha \omega^a$ and $\beta$ represent what are known as the shift and lapse of the time evolution vector. It is the lapse function $\beta$ which vanishes at the horizons. 

\subsection{Hamiltonian formulation}
We will now briefly review the Hamiltonian formalism for field theories on Kerr-like backgrounds. 
This will serve to familiarize ourselves with the concepts and notations needed to address 
constrained field theories. 
As mentioned above, time evolution is taken to be along $\xi^a$\,.
This ensures that the fields evolve in time while the background on which they are 
defined remains fixed. More specific to the Hamiltonian formalism, it ensures that 
Hamilton's equations take their usual form without any modification of the (covariant) 
definition of the Poisson bracket. 
Thus for any field $\Phi_A$\,, we have 
\begin{equation}
\dot{\Phi}_A := \pounds_{\chi} \Phi_A - \pounds_{\alpha \omega} \Phi_A = \pounds_{\xi} \Phi_A \,.
\label{gen.dot}
\end{equation}

The action functional for a field $\Phi_A$ is given by the time 
integral of the Lagrangian $L$, or equivalently the integral of 
the Lagrangian density ${\cal L}$ over the four volume,
\begin{equation}
S[\Phi_A] = \int dt\, L  
\equiv \int dt \int \limits_{\Sigma} \beta dV_x\,\, ~ {\cal L}(\Phi_A(x) , \nabla_a \Phi_A(x)) \, ,
\label{gen.act}
\end{equation}
where $dV_x\,\,$ is the volume element on $\Sigma\,$ 
and ${\cal L}(\Phi_A(x) , \nabla_a \Phi_A(x))$ is the Lagrangian density. 
Denoting the spacetime volume element in the orthonormal basis by $\epsilon_{abcd}$ and the spatial volume element of the hypersurface by $^{(3)}\epsilon_{bcd}$, we have
\begin{equation}
\chi^a \epsilon_{a b c d} = \beta \,^{(3)} \epsilon_{b c d} = \xi^a \epsilon_{a b c d} \, . 
\end{equation}
Thus the projected volume element has the correct form even though time evolution takes place
along $\xi^a$\, while it is $\chi^a$ which is orthogonal to $\Sigma\,.$ 

The canonically conjugate momenta $\Pi^A$ are defined as
\begin{equation}
\Pi^A(\vec{x}, t)  = \frac{\delta L}{\delta \dot{\Phi}_A(\vec{x}, t)}  \, ,
\label{gen.mom}
\end{equation}
where the functional derivative is an `equal-time' derivative evaluated on the hypersurface $\Sigma\,,$ 
\begin{equation}
\frac{\delta\Phi_A(\vec{x}, t)}{\delta\Phi_B(\vec{y}, t)} = \delta^B_A\, \delta(x, y) = 
\frac{\delta\dot\Phi_A(\vec{x}, t)}{\delta\dot\Phi_B(\vec{y}, t)}\, .
\label{gen.var}
\end{equation}
The $\delta(x, y)$ in Eq.~(\ref{gen.var}) is a three-dimensional covariant delta function
on $\Sigma$\,, satisfying
\begin{equation}
\int\limits_\Sigma dV_y\, \delta(x,y) f(\vec{y}, t) = f(\vec{x}, t)\,.
\label{gen.del}
\end{equation}
We will sometimes write $\vec{x}$ or even $(\vec{x},t)$ as $x$\,, etc. as we have done above.

The canonical Hamiltonian now follows from the Legendre transform
\begin{equation}
H_C = \int \limits_{\Sigma} dV_x\,\, ~(\Pi^A \dot{\Phi}_A) -  L \,.
\label{gen.Ham}
\end{equation}
The Poisson bracket is defined on the hypersurface, which for two 
functionals $F(\Phi_A(x), \Pi^A(x))$ and $G(\Phi_A(x), \Pi^A(x))$ of the fields and their momenta
is defined as
\begin{equation}
\left[F ,G\right]_P = \int dV_z \left[\frac{\delta F}{\delta \Phi_A(z)} \frac{\delta G}{\delta \Pi^A(z)} 
- \frac{\delta G}{\delta \Phi_A(z)} \frac{\delta F}{\delta \Pi^A(z)} \right]\,. \label{gen.PB}
\end{equation}
This definition provides the canonical Poisson brackets between the fields and their momenta, which follows from setting  
$F = \Phi_A(\vec{x}, t)$ and $G = \Pi^B(\vec{y}, t)$
\begin{equation}
\left[\Phi_A(\vec{x}, t), \Pi^B(\vec{y}, t) \right]_P = \delta_{A}^{B} \delta(x,y)\,. 
\label{gen.can}
\end{equation}
The Poisson bracket of any function or functional of the fields and momenta with the Hamiltonian provides its time evolution
\begin{equation}
\dot{F} = \left[F, H_C \right]_P\,.
 \end{equation}
This Hamiltonian provides a complete description of the dynamics of the system only if all 
the velocities are uniquely mapped into momenta through Eq.~(\ref{gen.mom}). This of course 
is not the case for constrained field theories. 
The constraints and dynamics of such theories can be determined from the Dirac-Bergmann formalism, 
which has been treated extensively in many excellent textbooks and 
reviews~\cite{Dirac-lect-1964, Hanson:1976cn, Henneaux:1992ig}.
In this formalism, constraints are classified into two types, first class and second class. 
Second class constraints can always be eliminated by using Dirac brackets while first class
constraints, apart from pathological counterexamples, generate gauge transformations.
All constrained field theories of interest to us, at least in this paper, are thus gauge field theories. 
We will find that the constraints include non-vanishing contributions from the fields on the horizon 
of the spacetime, thereby modifying the familiar constraints of theories on backgrounds 
without boundaries.  

In the following, we will demonstrate this by considering the Maxwell field.

\section{The Maxwell field}\label{Max}
The action for the Maxwell field is given by
\begin{equation}
S_{EM} =  \int dV_4 \left(-\tfrac{1}{4} F_{a b} F_{c d} g^{a c} g^{b d}\right) \, ,
\label{max.act}
\end{equation}
where $dV_4$ is the four dimensional volume form on the manifold $\Sigma\times\mathbb{R}$\,, and 
$F_{a b} = 2 \partial_{[a} A_{b]}$\,. Defining $e_a = - \beta^{-1} \chi^c F_{c a}$ and 
$f_{a b} = F_{c d} h^c_a h^d_b\,,$  and using the projection operator from Eq.~(\ref{gen.met})
 we can rewrite this action as
\begin{equation}
S_{EM} = - \int dt \int \limits_{\Sigma} dV_x\,\, \frac{\beta}{4} \left[f_{a b} f^{a b} - 2 e_{a} e^{a} \right] \, .
\label{H.Lag}
\end{equation}
From Eq.~(\ref{gen.dot}), we have 
\begin{align}
\dot A_b \equiv \pounds_{\xi} A_b  &= \pounds_{\chi}A_b - \alpha \pounds_{\omega}A_b - \left(A_a \omega^a \right) \nabla_b \alpha\, \notag \\
&=\chi^a F_{a b} + \nabla_b (A_a \xi^a) - \alpha \omega^a F_{a b} \,, 
\label{H.adot}
\end{align}
and defining 
	{$\phi = A_a \xi^{a}$},
	we have for the projection $a_b\,,$
\begin{equation}
\dot a_b = -\beta e_b + \mathcal{D}_b \phi + \alpha f_{b a} \omega^a \,.
\label{H.elf}
\end{equation}

It follows from Eq.~(\ref{H.adot}) that $\xi^a \dot{A}_a$ has its usual expression since $\chi^a - \alpha \omega^a = \xi^a$. Hence $\dot{\phi}$ is absent from the action in Eq.~(\ref{H.Lag}), leading to its conjugate momentum being a constraint,
\begin{equation}
 \pi^{\phi} = \frac{\partial L_{E M}}{\partial \dot{\phi}} = 0 \,.
\label{H.con}
\end{equation}
The momenta conjugate to the $a_b$ are given by
%
\begin{equation}
\pi^b = \frac{\partial L_{E M}}{\partial \dot{a}_b} = - e^b.
\label{H.mom}
\end{equation} 
The canonical Hamiltonian follows from the usual definition
\begin{align} 
H_C &= \int \limits_{\Sigma} dV_x\, \, \left(\pi^b \dot{a}_b\right) -  L \notag \\
&= \int \limits_{\Sigma} dV_x\, \, \left[\beta\left(\frac{1}{2} \pi^b \pi_b + \frac{1}{4} f_{a b} f^{a b}\right) + \pi^b \mathcal{D}_b \phi  + \alpha\pi^b f_{b a} \omega^a\right]\,.
\end{align}
The Hamiltonian comprises of the usual energy density along with an energy current $\alpha \pi^b f_{ba}\omega^a$ due to the non-vanishing shift vector of the background. This is a known current
which has been found elsewhere in considerations of the Maxwell field on foliated backgrounds involving a non-vanishing shift vector~\cite{MacDonald:1982zz,Bojowald:2010qpa}.
By including the constraint of Eq.~(\ref{H.con}) to the canonical Hamiltonian, a new 
Hamiltonian is defined,
\begin{equation}
H_0  = \int \limits_{\Sigma} dV_x\, \, \left[\beta \left(\frac{1}{2} \pi^b \pi_b + \frac{1}{4} f_{a b} f^{a b}\right) + \pi^b \mathcal{D}_b \phi + \alpha \pi^b f_{b a} \omega^a + v_{\phi} \pi^{\phi} \right] \, ,
\label{H.pri}
\end{equation}
where $v_{\phi}$ is an undetermined multiplier. The canonical Poisson brackets of
Eq.~(\ref{gen.can}) are in this case
\begin{align}
\left[\phi(x), \pi^{\phi}(y) \right]_P & = \delta(x,y) \notag\\
\left[ a_a(x) , \pi^b(y) \right]_P & = \delta^{b}_{a}\delta(x,y) \,.
\label{H.can}
\end{align}
%
%
\subsection{The Dirac-Bergmann formalism} \label{MDBA}
We will now determine all additional constraints of the theory and construct the unconstrained Hamiltonian through the 
Dirac-Bergmann formalism. This requires the Poisson brackets of the constraints with the Hamiltonian, which we will 
always evaluate with the help of smearing functions. As we will see, these smearing functions come from
the same space as the functions of gauge transformations, i.e. the dual space of the space 
of the gauge generators~\cite{Benguria:1976in}. Therefore we will not assume that these smearing 
functions vanish at the horizons, but only that they are regular there. 
We need to check that the constraints of the theory are obeyed at all times, 
or in other words, $\dot \pi^{\phi} \approx 0 \,.$ This is done using the Poisson bracket between
$\pi^\phi$ and the Hamiltonian, and with the help of a smearing function $\epsilon$ as follows,
\begin{align}
\int \limits_{\Sigma} dV_y\, \epsilon(y) \dot\pi^\phi(y) &= \int \limits_{\Sigma} dV_y\, 
\epsilon (y) \left[\pi^{\phi}(y) , H_0 \right]_P  \,\notag \\
&= \int \limits_{\Sigma} dV_y\, 
\epsilon (y) \left[\pi^{\phi}(y) , \int \limits_{\Sigma} dV_x\, \pi^b(x) 
\mathcal{D}^x_b \phi(x) \right]_P \notag \\
&=   \oint \limits_{\partial \Sigma} da_y \, \epsilon(y)  n^y_b \pi^b(y) + 
\int  \limits_{\Sigma} dV_y\, \, \epsilon(y) \left( \mathcal{D}^y_b \pi^b(y)\right)\,.
\label{U.PB1}
\end{align}
In deriving the above result we used the canonical Poisson brackets given in Eq.~(\ref{H.can}) 
and an integration by parts. The $n_b$ involved in the surface integral over the horizons of 
the spacetime is the `unit normal' to the surface of the horizon satisfying $n_b n^b = 1$ and
pointing into the region where $\chi^a$ is timelike. 
We are particularly interested in the case where $\partial\Sigma$ consists of an outer cosmological horizon 
	and an inner black hole horizon.
Here and for the rest of the paper, we have allowed the smearing functions and its derivatives 
to be non-vanishing but regular at the horizons. Then by the Schwarz inequality we have
\begin{equation}
\left| n_b \pi^b \right|  \leq \, \sqrt{\left|n_b n^b\right| \, 
	\left|\pi_b \pi^b\right|}  \,.
\label{U.SI}
\end{equation}
$n_b n^b = 1$ and $\pi_b\pi^b = e_b e^b$ appears in the energy momentum tensor (more precisely
in invariant scalars such as $T^{ab}T_{ab}$), and therefore may not
diverge at the horizon.  
Thus when the smearing function $\epsilon$ is regular at the horizon, 
the surface integral provides a finite contribution from $\partial\Sigma$\,. 
Hence setting the right hand side of Eq.~(\ref{U.PB1}) to weakly vanish 
produces the constraint
\begin{equation}
\int \limits_{\Sigma} dV_x \epsilon(x)\Omega_2(x) = \int \limits_{\Sigma} dV_x  \epsilon(x)\mathcal{D}^x_b \pi^b(x) + \oint \limits_{\partial \Sigma} da_x \epsilon(x) n^x_b  \pi^b(x)  \approx 0 \,,
\label{U.con2int}
\end{equation}
In the absence of a surface integral, as for spacetimes without bounadries, we would extract the smearing function 
$\epsilon(x)$ and write the constraint as a weakly vanishing distribution on $\Sigma\,,$ 
with the understanding 
that manipulations involving the constraint requires a smearing function 
and integration over the volume. In the present case, we will express the constraint appearing in 
Eq.~(\ref{U.con2int}) as 
\begin{equation}
\Omega_2 =  \mathcal{D}_b \pi^b + n_b  \pi^b\Big\vert_{\cal{H}} \approx 0 \,.
\label{U.con2}
\end{equation}
The notation $\vert_{\cal{H}}$ symbolizes that this term must be integrated with respect to the area 
element at the horizons. In other words, while the usual Gauss' Law constraint holds for all points of 
$\Sigma\,,$ the additional surface contribution in Eq.~(\ref{U.con2}) must be considered for all points 
at the horizon $\partial \Sigma$. 
This is a key result of our paper.

It is now straightforward to verify that $\dot{\Omega}_2 = \left[\Omega_2(x) , H_0 \right]_P 
\approx 0\,,$ which reveals that there are no further constraints.
Thus the full Hamiltonian is given by
\begin{align}
H_T = \int\limits_{\Sigma} dV_x\, \left[ \beta \left( \frac{1}{4} f_{a b} f^{a b} + \frac{1}{2} \pi_a \pi^a\right) \right. &\left. + v_1\left( \mathcal{D}_b \pi^b \right) + \pi^b \mathcal{D}_b \phi + \alpha \pi^b f_{b a} \omega^a +  v_{\phi} \pi^{\phi}\phantom{\frac{1}{2}} \right] \notag\\ 
& \qquad \qquad + \int\limits_{\partial \Sigma} v_1 n_b \pi^b\,.
\label{U.Htot1}
\end{align}
The multipliers $v_1$ and $v_\phi$ may be determined by examining the equations of motion. 
The evolution of $\phi$ is given by
\begin{align}
\int \limits_{\Sigma} dV_y\, \epsilon(y) \dot \phi(y) &= \int \limits_{\Sigma} dV_y\, 
\epsilon(y)\left[  \phi(y) , H_T \right]_P \notag \\
&=  \int \limits_{\Sigma} dV_y\, \epsilon(y) \int \limits_{\Sigma} dV_x\, v_{\phi}(x) 
\left[ \phi(y) , \pi^{\phi}(x) \right]_P \notag \\
&=  \int \limits_{\Sigma} dV_y\, \epsilon (y) v_{\phi}(y) \, ,
\end{align}
which tells us that we can set $\dot \phi = v_{\phi}$. The evolution of $a_b$ can also 
be determined in a similar manner,
\begin{align}
\int \limits_{\Sigma} & dV_y\,\, \epsilon(y) \dot{a}_b(y) = \int \limits_{\Sigma} dV_y\,\, \left[ \epsilon(y) a_b(y) , H_T \right]_P \notag \\
&=  \int \limits_{\Sigma} dV_y\, \epsilon(y) \left[ \beta(y) \pi_b(y) + \mathcal{D}^y_b \phi(y) + \alpha f_{b a} \omega^a - \mathcal{D}^y_b v_1(y) \right] \, .
\label{U.vpt}
\end{align}
Comparing this with Eq.~(\ref{H.elf}), we deduce that $\mathcal{D}_b v_1 =  0$. 
With this choice,  Eq.~(\ref{U.vpt}) produces
\begin{equation}
\dot a_b = \beta\pi_b + \mathcal{D}_b \phi + \alpha f_{b a}\omega^a \, ,
\end{equation}
and this can result by simply setting $v_1 = 0$. While this choice is not unique, we would always have $v_1 \Omega_2 = \pi^b \mathcal{D}_b v_1 = 0$.  Hence the total Hamiltonian takes the form
\begin{equation}
H_T = \int\limits_{\Sigma} dV_x\, \left[\beta \left(\frac{1}{4} f_{a b} f^{a b} + \frac{1}{2} \pi_a \pi^a\right) + \pi^b \mathcal{D}_b \phi + \alpha \pi^b f_{b a}\omega^a + \dot{\phi} \pi^{\phi} \right] \,.
\label{U.fHam}
\end{equation}
The two first class constraints generate gauge transformations on the fields. 
By evaluating the Poisson bracket of the fields $\phi$ and  $a_b$ with the general 
linear combination of the constraints $\rho = \epsilon_1 \pi^{\phi} + \epsilon_2 \Omega_2$, we find that
\begin{align}
	\delta \phi(y) &= \left[ \phi(y) , \rho(x) \right]_P = \epsilon_1 (y) \notag \\
	\delta a_b(y) &= \left[ a_b(y) , \rho (x) \right]_P = -\mathcal{D}^y_b \epsilon_2(y)\, .
	\label{GT1}
\end{align}
The gauge transformations which leave the Lagrangian in Eq.~(\ref{max.act}) invariant are $\delta A_{b} = \nabla_{b} \epsilon$. 
By projecting this expression using Eq.~(\ref{gen.met}) we have
\begin{align}
\delta \left(\phi + \alpha \omega^a a_a\right) = \pounds_{\chi} \epsilon \, , \qquad \delta a_b = \mathcal{D}_b \epsilon\,.
\label{GT2}
\end{align}
Eq.~(\ref{GT1}) is equivalent to Eq.~(\ref{GT2}), provided we identify $ \epsilon_1 (y) = \pounds_{\xi} \epsilon(y)$ 
and $\epsilon_2 (y) = -\epsilon(y)$. Without the horizon term in the constraint, the 
gauge transformations will clearly not have the usual form unless $\epsilon$ is assumed to vanish on the horizon.
%

\subsection{Gauge fixing}\label{sec.gf}

We will look at this theory in two different gauges -- the 
radiation gauge and the axial gauge. In the radiation gauge the 
full set of constraints are $\Omega_i \approx 0\,,$ with
\begin{align}
\Omega_1 & = \pi^{\phi} \notag \\
	\Omega_2 & =  \mathcal{D}_a \pi^a
	+  n_a \pi^a \Big\vert_{\cal{H}} \notag\\
	\Omega_3 & = \phi \notag \\
	\Omega_4 & = \mathcal{D}^b\left(\beta a_b\right) \, .
	\label{gf.con}
\end{align}
The first two are the gauge constraints of the theory 
already found in Eq.~(\ref{H.con}) and Eq.~(\ref{U.con2}), while 
$\Omega_3$ and $\Omega_4$ are the gauge-fixing functions. We call this the radiation 
gauge because that is what it reduces to in flat space.
The full set of constraints 
is second-class, requiring the construction of Dirac brackets.

The non-vanishing Poisson brackets of the constraints in Eq.~(\ref{gf.con}) are 
\begin{align}
	\left[\Omega_1 (x), \Omega_3 (y) \right]_P & = - \delta(x,y)\,, \notag \\
	\left[\Omega_2 (x), \Omega_4 (y) \right]_P & =  
	 \mathcal{D}_{a} \left(\beta\mathcal{D}^{a}\delta(x,y) \right)\,. 
	\label{gf.pb}
\end{align} 
The first Poisson bracket is the canonical relation given in Eq.~(\ref{H.can}). The 
second Poisson bracket is calculated as follows.
\begin{align}
	& \left[ \int \limits_{\Sigma} dV_x \, \eta(x) \Omega_2 (x) , 
	\int \limits_{\Sigma}dV_y\, \epsilon(y) \Omega_4(y) \right]_P \notag\\
&\qquad\qquad = \left[ \int \limits_{\Sigma} dV_x\, \eta(x) \mathcal{D}^x_a \pi^a(x)
	 + \oint \limits_{\partial \Sigma}\, \eta(x) n^x_a \pi^a(x)  , 
	\int \limits_{\Sigma} dV_y\, \epsilon(y)\mathcal{D}_y^b\left(\beta(y) a_b(y)\right)\right]_P \notag\\
	 & \qquad \qquad  = \left[ \int \limits_{\Sigma} dV_x\,  \left(\mathcal{D}^x_a 
	 \eta(x) \right)  \pi^a(x)\,, \int \limits_{\Sigma} dV_y\, \beta(y) \left(\mathcal{D}_y^b\epsilon(y) \right) a_b(y) \right]_P \notag \\
	& \qquad \qquad  = - \int \limits_{\Sigma} dV_y\, \beta(y) \left(\mathcal{D}^y_a
	\eta(y)\right) \left(\mathcal{D}_y^a\epsilon(y) \right) \notag \\
	& \qquad \qquad  = \oint \limits_{\partial \Sigma} da_y \epsilon(y) n_y^a \beta(y) \left(\mathcal{D}^y_a\eta(y)\right) + \int \limits_{\Sigma} dV_y\, \epsilon(y) \mathcal{D}_y^a\left(\beta(y) \mathcal{D}^y_a\eta(y)\right).
	\label{gf.pb2}
\end{align}
By using Schwarz's inequality on the surface term in the last line above, we get
\begin{align}
	\left| n^a \beta D_a \left(\eta\right) \right|^2 & \leq \left|n^a n_a \right| \beta^2 \left|\left(D_a\eta\right) \left(D^a\eta\right)\right| \notag\\
	& = \beta^2 \left(D_a\eta\right) \left(D^a\eta\right) \, .
	\label{gf.srf}
\end{align}
Due to the presence of $\beta^2$, the surface integral vanishes and only the second term of Eq.~(\ref{gf.pb2}) contributes. The Poisson bracket in Eq.~(\ref{gf.pb2}) can thus be written as 
\begin{equation}
\left[ \int \limits_{\Sigma} dV_x\, \eta(x) \Omega_2 (x) ,\int \limits_{\Sigma} dV_y\, \epsilon(y) \Omega_4(y) \right]_P =  \int \limits_{\Sigma} dV_y\,  \epsilon(y) \int \limits_{\Sigma} dV_x\, \eta(x) \left[ \mathcal{D}_y^a\left(\beta(y)\mathcal{D}^y_a  
	\delta (x,y)\right) \right]\,,
\label{gf.pbr}
\end{equation}
which corresponds to the result given in Eq.~(\ref{gf.pb}).
The matrix of the Poisson brackets between these constraints have 
a non-vanishing determinant and is invertible. This matrix, 
$C_{\alpha\beta}\left(x,y \right) = \left[\Omega_\alpha (x), 
\Omega_\beta(y)\right]_P\,,$ is given by
\begin{equation}
	C (x,y) = 
	\begin{pmatrix} 
		0 & 0 & -\delta(x,y)\ & 0 \\
		0 & 0 & 0 & \mathcal{D}_a\left(\beta\mathcal{D}^a\delta(x,y)\right) \\
		\delta(x,y) & 0 & 0 & 0 \\
		0 & - \mathcal{D}_a\left(\beta\mathcal{D}^a\delta(x,y)\right) & 0 & 0
	\end{pmatrix} \, .
	\label{gf.dirm}
\end{equation}
The dynamics of the gauge fixed theory is determined through Dirac brackets, whose definition requires the inverse 
of the matrix given in Eq.~(\ref{gf.dirm}). The Dirac brackets of the theory for two dynamical 
entities $A$ and $B$ (which may be functions or functionals on phase space) is defined as
\begin{equation}
	\left[A\,, \,B \right]_{D} = \left[A\,, \,B \right]_{P} - \int \limits_{\Sigma} dV_u \int \limits_{\Sigma} dV_v 
	\left[A\,, \,\Omega_\alpha(u)\right]_P C^{-1}_{\alpha\beta}\left(u, v\right) \left[\Omega_\beta(v)\,, \,B\right]_P\,.
	\label{gf.dir}
\end{equation}  
Thus we need to find the inverse of the operator
$\mathcal{D}_a
		\left(\beta\mathcal{D}^a\right)$.
Let us formally write the inverse as $G(x, y)\,,$ i.e. 
\begin{equation}
	\mathcal{D}_a\left(\beta\mathcal{D}^a G\left(x, y\right)\right) = -\delta\left(x,y\right) \, ,
	\label{U.gfs}
\end{equation}
for some scalar function $G\left(x,y\right)$. This is the {\em time-independent} and axisymmetric Green's function 
for the spacetime Laplacian operator as can be easily verified by projecting it on the hypersurface. Thus the inverse of the matrix in Eq.~(\ref{gf.dirm}), $C^{-1}_{\alpha\beta}(x,y)$, is now given by 
\begin{equation}
	C^{-1}(x,y) = 
	\begin{pmatrix}
		0 & 0 & \delta(x,y) & 0\\
		0 & 0 & 0 & G\left(x,y\right) \\
		-\delta(x,y) & 0 & 0 & 0 \\
		0 & - G\left(x,y \right) & 0 & 0
	\end{pmatrix} \, .
	\label{gf.cin}
\end{equation}
Using Eq.~(\ref{gf.cin}) in Eq.~(\ref{gf.dir}) we find that the non-vanishing Dirac brackets are
\begin{align}
	\left[a_a(x),\pi^b(y)\right]_{D} & = \delta(x,y)\delta_a^b - \mathcal{D}_a^x\left(\beta(y)\mathcal{D}^b_y G\left(x,y \right)
	\right)\,. 
\label{gf.cdir}
\end{align}
The Green function involved in Eq.~(\ref{gf.cdir}) has a known closed 
form expression outside the ergosphere on the Kerr background~\cite{Ottewill:2012aj}. 
For the electromagnetic field on the Schwarzschild background, it is known 
that the Dirac bracket in the radiation gauge reduces to the Poisson bracket when 
either $\pi^b$ or $a_a$ is at the horizon~\cite{Fernandes:2016sue}. 
This does not occur if we use a modified radiation gauge which involves a surface 
term at the horizon. Since expressions for the scalar Green function valid up to the 
horizon of a Kerr black hole are not known, such an analysis cannot be performed on 
axisymmetric spacetimes. We will however further elaborate on the implications of 
a radiation gauge with a surface term at the horizons in the discussion 
section of this paper.

Given that our background is axisymmetric, we will now further consider the axial gauge. 
Our consideration of the axial gauge will generalize the treatment provided in~\cite{Hanson:1976cn} 
about flat spacetime. We adopt the basis $\{\phi^a \,,\mu^a \, , \nu^a \}$ described in 
Sec.~\ref{geom} and will consider Eq.~(\ref{gf2.proj2}) in the following equations.  
While it is possible to identify and select the axial direction by an appropriate 
choice of coordinates in the $\mu$-$\nu$ plane, we can work with any linear combination 
of $\mu^a$ and $\nu^a$. Let us fix a `generalized axial gauge' by setting to zero 
the component of $a_a$ along $\mu^a$. We then have the following set of constraints 
in this gauge~\cite{Hanson:1976cn}:
\begin{align}
\Omega_1 & = \pi^{\phi} \notag \\
	\Omega_2 & =  \mathcal{D}_b \pi^b
	+ n_b \pi^b \Big\vert_{\cal{H}}  \notag\\
	\Omega_3 & = \mu^a a_{a} \notag \\
	\Omega_4 & = \mu^a\mathcal{D}_{a}\phi+\beta \mu^a \pi_{a} + \alpha \mu^af_{ac}\omega^c\,.
	\label{gf2.gf}
\end{align}

The constraints have the non-vanishing Poisson brackets  
\begin{align}
	\left[\Omega_1 (x), \Omega_4 (y) \right]_P & = -\mu^a(y) \mathcal{D}^{y}_{a}\delta(x,y) =\left[\Omega_4 (x), \Omega_1 (y) \right]_P\,, \notag \\
	\left[\Omega_2 (x), \Omega_3 (y) \right]_P & = \mu^a(y)\mathcal{D}^{y}_{a}\delta(x,y) =\left[\Omega_3 (x), \Omega_2 (y) \right]_P\,,\notag\\
	\left[\Omega_3 (x), \Omega_4 (y) \right]_P & = \beta(y)\delta(x,y)\,.
	\label{gf2.pb}
\end{align} 
The bracket $\left[ \Omega_2(x) \, , \Omega_4(y) \right]_P$ vanishes because we have assumed that there is no torsion,
\begin{align}
&\left[ \int \limits_{\Sigma} dV_x\, \eta(x) \Omega_2 (x) , \int \limits_{\Sigma} dV_y\, \epsilon(y) \Omega_4(y) \right]_P \notag\\
&\quad = \left[ \int \limits_{\Sigma} dV_x\, \eta(x) \mathcal{D}^x_b \pi^b(x)
	+ \oint \limits_{\partial \Sigma} da_x\, \eta(x) n^x_b \pi^b(x)  , \int \limits_{\Sigma} dV_y\, \epsilon(y) \left(\mu^a\mathcal{D}_{a}\phi+\beta \mu^a \pi_{a} + \alpha \mu^af_{ac}\omega^c\right)(y) \right]_P \notag\\
&\qquad \qquad  =  \int \limits_{\Sigma} dV_x\, \mathcal{D}_a^x (\eta(x))  \mathcal{D}_b^x \left(\alpha(x) \epsilon(x) \left(\mu^b(x) \omega^a(x) - \mu^a(x) \omega^b(x)\right)\right) \notag\\
&\qquad  \qquad \qquad \qquad  + \oint \limits_{\partial \Sigma} da_x \mathcal{D}_a^x (\eta(x))  n_b^x \left(\alpha(x) \epsilon(x) \left(\mu^b(x) \omega^a(x)\right)\right) \notag\\
&\qquad \qquad = - \int \limits_{\Sigma} dV_x\,  \alpha(x) \epsilon(x) \left(\mu^b(x) \omega^a(x) - \mu^a(x) \omega^b(x)\right) \mathcal{D}_b^x \mathcal{D}_a^x (\eta(x)) \notag\\
&\qquad \qquad = 0\,.
\label{gf.pbr24}
\end{align}
%
%
%
The Poisson brackets of (\ref{gf2.pb}) lead to the following matrix,
\begin{equation}
	C(x,y) = 
	\begin{pmatrix} 
		0 & 0 & 0 & -\mu^a(y) \mathcal{D}^{y}_{a}\delta(x,y)\\\
		0 & 0 & \mu^a(y) \mathcal{D}^{y}_{a}\delta(x,y) & 0 \\
		0 & \mu^a(y) \mathcal{D}^{y}_{a}\delta(x,y) & 0 & \beta(y)\delta(x,y)\\
		-\mu^a(y) \mathcal{D}^{y}_{a}\delta(x,y) & 0 & -\beta(y)\delta(x,y) & 0
	\end{pmatrix} \, .
	\label{gf2.mat}
\end{equation}
The inverse of this matrix is needed for the Dirac brackets. Let us write it as
\begin{equation}
	C^{-1}(x,y) = 
	\begin{pmatrix}
		0 & -p(x,y) & 0 & q(x,y)\\
		p(x,y) & 0 & -q(x,y) & 0 \\
		0 & -q(x,y) & 0 & 0 \\
		q(x,y) & 0 & 0 & 0
	\end{pmatrix} \, ,
	\label{gf2.inv}
\end{equation}
where $p(x,y)$ and $q(x,y)$ are two functions which may be found by 
evaluating $\int dV_z C(x,z)C^{-1}(z,y) = \delta(x,y)$\,. 
We find that these functions must satisfy 
\begin{align}
\mu^a(y) \mathcal{D}^{y}_{a}q(x,y)&= -\delta(x,y) \label{gf.q}\\
\mu^a(y)\mathcal{D}^{y}_{a}p(x,y)&= - \beta(y) q(x,y) \label{gf.p}
\end{align} 

The expressions for $p$ and $q$ on the asymptotically flat Kerr background in Boyer-Lindquist coordinates are derived in Appendix~\ref{A}. Since Eq.~(\ref{gf.q}) and Eq.~(\ref{gf.p}) involve first order differential equations, their solutions will also exist on other Kerr-like backgrounds.
Using the matrix of Eq.~(\ref{gf2.inv}) and the constraints given in Eq.~(\ref{gf2.gf}), we derive the following non-vanishing Dirac brackets for the fields,
\begin{align}
\left[\phi(x),a_b(y) \right]_{D} & = \mu_b(y)\beta(y) q(x,y) + \mathcal{D}^{y}_{b}p(x,y) \, ,\label{gf2.one}\\
\left[\phi(x),\pi^b(y) \right]_{D} & = \mathcal{D}_a^y \left( \alpha(y) q(x,y) (\mu^a(y) \omega^b(y)  - \mu^b(y) \omega^a(y))\right)\notag\\
&\qquad \qquad + n_a^y \alpha(y) q(x,y) (\mu^a(y) \omega^b(y))\vert_{\cal{H}} \, , \label{gf2.two}\\ 
\left[a_b(x),\pi^c(y)\right]_{D} & =  \delta^c_b \delta(x,y)+\mu^c(y)\mathcal{D}^{y}_{b}q(x,y)\,.
\label{gf2.dir}
\end{align}
The Dirac bracket in Eq.~(\ref{gf2.two}), which also involves contributions from the horizons of the spacetime, is not present in flat space results involving the axial gauge. It appears here due to the non-vanishing shift vector of the Kerr background. The derivation of the bracket is provided in Appendix~\ref{app.der}.

Use of Dirac brackets ensures that all brackets involving $\mu^a a_{a}$\,, or the other constraints 
in Eq.~(\ref{gf2.gf}), identically vanish. The Hamiltonian in Eq.~(\ref{U.fHam}) becomes, after the constraints of Eq.~(\ref{gf2.gf}) have been imposed, 
\begin{align}
H_T &= \int\limits_{\Sigma} dV_x\, \left[\beta \left(\frac{1}{4} f_{a b} f^{a b} + \frac{1}{2} (f^{-2}\omega^a \omega_b + \nu^a \nu_b)\pi_{a} \pi^{b}\right) + (f^{-2}\omega^a \omega_b + \nu^a \nu_b)\pi^{b} \mathcal{D}_{a} \phi \right. \notag\\
& \left.  \qquad   \qquad \qquad  \qquad \qquad -\frac{1}{2}\beta \mu^a \pi_a \mu^b \pi_b + \alpha  \nu_b \nu^c \pi^{b} f_{c a} \omega^a \right] \,.
\end{align}
%
%
\section{Charges and currents} \label{CC}
The modification of Gauss law by horizon terms has interesting consequences --
in particular, it ties in nicely with the membrane paradigm  as we shall see below.
The electric and magnetic fields appearing in the analysis of constraints are in general from 
external sources, we do not assume that they share the symmetries of 
the background. We first note that Maxwell's equations resulting from the Hamiltonian, 
derived in Appendix~\ref{App.C}, are given by
\begin{align}
\pounds_{\chi} \pi^{b} &= \mathcal{D}_a(\beta f^{ab}) +\alpha \omega^b \left(\mathcal{D}_a\pi^a + n_a \pi^a \Big \vert_{\cal{H}}\right) \approx \mathcal{D}_a(\beta f^{ab})  \, , \label{Ham.el2}\\
\pounds_{\chi} f_{ab} &= 2 \mathcal{D}_{[a} \beta \pi_{b]}  \label{Ham.mag2} \, .
\end{align}
The equations of motion involve a term proportional to the Gauss law constraint,
which does not affect the dynamics of $\pi^b$ since the constraint vanishes weakly. 
This is nothing unusual, Hamiltonian equations of motion hold up to constraints. 
Thus we find that for electromagnetism in black hole spacetimes, while the Gauss 
law constraint is modified by surface terms at the horizons, the dynamical Maxwell 
equations are not. The Gauss law constraint 
can be used to determine the charge contained in a given region of the spacetime. 
By considering a region from the black hole horizon $\mathcal{H}$ to an outer 
(spacelike) boundary $\partial \Sigma_B$, we have
\begin{align}
	Q_B &= \int \limits_{\Sigma_B} dV_x \, \Omega_2(x) \notag\\
&=\int \limits_{\Sigma_B} dV_x \, \mathcal{D}_b^x \pi^b(x) + {\overline{\oint \limits_{\cal{H}}}} da_x  \,n^x_b \pi^b(x) \notag\\
	&= - \oint \limits_{\partial \Sigma_B} da_x \,n^x_b \pi^b(x) - \oint \limits_{\cal{H}} da_x  \,n^x_b \pi^b(x) 
	+ {\overline{\oint \limits_{\cal{H}}}} da_x  \,n^x_b \pi^b(x) \notag\\
	&= - \oint \limits_{\partial \Sigma_B} da_x  \,n^x_b \pi^b(x)\,.
	\label{con.ch}
\end{align}
We have introduced the notation $\displaystyle{\overline \oint}$ to 
indicate any surface integral which arises from the horizon term in 
the Gauss law constraint. The surface integrals have their usual meaning and the 
notation is merely used to keep track of contributions from the surface terms in 
Gauss law. We see that for an observer outside the horizon, the enclosed charge 
is determined by the usual expression of the electric flux across $\partial \Sigma_B$. 
The surface term in the Gauss law constraint only contributes a surface integral at 
$\mathcal{H}$ and does not provide a term at $\partial \Sigma_B$. However, now let us 
shrink the surface to the horizon, $\partial \Sigma_B \to \mathcal{H}$. If we do the 
same calculation now, we will get an additional contribution  
${\overline{\oint \limits_{\cal{H}}}} da_x  \,n^x_b \pi^b(x)$ from the surface term 
in the Gauss law constraint, resulting in a vanishing charge at the black hole horizon
\begin{equation}
Q_{\cal{H}} = 0\,.
\label{ch.bh}
\end{equation}
Thus the non-vanishing electric flux outside the horizon is seen to vanish by an 
observer at the horizon. Clearly it is the surface term in the constraint which 
causes the total charge to vanish for an observer on the horizon. 
This suggests that the surface term contribution in the 
Gauss law constraint corresponds to an induced charge on the horizon of the black hole.
We can define the induced surface charge density $\sigma$ at the black hole horizon by
\begin{equation}
n^x_b \pi^b(x) \Big \vert_{\cal{H}} = \sigma(x) \Big \vert_{\cal{H}} \,.
\label{el.prp}
\end{equation}
As the spacetime is rotating, we can also identify a surface current density on the horizon. 
Contracting Eq.~(\ref{Ham.el2}) with the unit normal at the horizon, we find
\begin{equation} 
\pounds_{\chi} \sigma = \mathcal{D}_a\left(\beta f^{ab}n_b\right) \,.
\end{equation}
This equation represents the expression for charge conservation on Kerr spacetimes
\begin{equation}
\pounds_{\chi} \sigma + \mathcal{D}_aj^a = 0\,,
\end{equation}
provided we have 
\begin{equation}
\beta n_a f^{ab} = j^b\,, 
\label{cc.indI}
\end{equation} 
as the induced current on the black hole horizon. Since the current is parallel to 
the horizon, $\mathcal{D}_a j^a$ is the two-dimensional divergence on the surface of the horizon. 
If we define the magnetic field $B_c$ as
\begin{equation}
\beta f^{ab} = - \epsilon^{abc}B_c \,,
\end{equation}
then Eq.~(\ref{cc.indI}) is satisfied given the following expression for the parallel 
components of the magnetic field $B^a_{\parallel}$
\begin{equation}
B^a_{\parallel} \Big \vert_{\cal{H}} = \epsilon^{abc}j_b n_c \Big \vert_{\cal{H}}
\label{mag.pll}
\end{equation}
The above treatment extends to backgrounds with an outer cosmological horizon. In this case, 
by integrating the Gauss law over the entire hypersurface, whose inner boundary is the 
black hole horizon $\mathcal{H}_1$ and outer boundary is the cosmological horizon $\mathcal{H}_2$, we have
\begin{align}
	Q &= \int \limits_{\Sigma} dV_x \, \Omega_2(x) \notag\\
&=\int \limits_{\Sigma} dV_x \, \mathcal{D}_b^x \pi^b(x) + {\overline{\oint \limits_{\cal{H}}}} da_x  \,n^x_b \pi^b(x) \notag\\
	&= - \oint \limits_{\mathcal{H}_2} da_x \,n^x_b \pi^b(x) - \oint \limits_{\mathcal{H}_1} da_x  \,n^x_b \pi^b(x) + {\overline{\oint \limits_{\mathcal{H}_2}}} da_x \,n^x_b \pi^b(x) + {\overline{\oint \limits_{\mathcal{H}_1}}} da_x  \,n^x_b \pi^b(x) \notag\\
	&= 0\,.
	\label{con.ch2}
\end{align} 
The surface charge density in Eq.~(\ref{el.prp}) and surface current density in 
Eq.~(\ref{mag.pll}) can now be defined on both $\mathcal{H}_1$ and $\mathcal{H}_2$.
We hence have the following situation. The Gauss law constraint on backgrounds with 
horizons is modified by surface terms at the horizons of the spacetime, which can be 
identified with induced surface charge densities defined locally on these horizons. 
These induced charges lead to a vanishing electric flux at the horizons and is related 
to the normal component of the electric field. On the other hand from Maxwell's equations, 
we can also determine induced induced surface current densities on the horizons of Kerr 
spacetimes. These current densities are related to parallel components of the magnetic 
field on the horizons. 

The induced charges and currents that we find on black hole horizons have been introduced 
before in the literature. It was noted in~\cite{Hanni:1973fn} that when an electric charge 
is lowered into a Schwarzschild black hole, the electric flux lines terminate on the horizon. 
This required the introduction of an induced surface charge density on the horizon, and the 
electric potential was calculated as the superposition of that due to the external
charge and that due to the induced charge.
This result was generalized to describe an induced surface current density on the horizon 
of a rotating black holes in an asymptotically flat spacetime in~\cite{Damour:1978cg, Znajek:1978}. 
The induced surface charges and currents can be described within the membrane paradigm as 
conditions on the electromagnetic fields on the membrane~\cite{MacDonald:1982zz,Thorne:1986iy}
as well as through a surface action for the electromagnetic field on the membrane~\cite{Parikh:1997ma}. 
The induced charges and currents on the horizon help describe the Blandford-Znajek 
mechanism~\cite{Blandford:1977ds}, a magnetic Penrose process which provides a model 
for the source of pulsars, quasars and active galactic nuclei~\cite{Granot:2015xba,Piran:2005qu}. 
Our result demonstrates that induced charges and currents on the horizon arise naturally 
as part of the general Gauss law constraint on black hole backgrounds. 
In the membrane paradigm, the induced charge density on the horizon appears as a consequence of
boundary conditions .
The vanishing electric flux at the horizons, following our treatment, 
could provide a means to investigate soft limits and their relation to gauge parameters at the 
horizon. In this regard, we note the proposal in~\cite{Hawking:2016msc}, where soft hairs were 
defined as charges on the future horizon of the black hole, considered as a `holographic plate', 
which are associated with non-vanishing large gauge transformations on the horizon. It will be 
interesting to investigate if such charges also result for the quantized electromagnetic 
field as a consequence of gauge parameters and constraints at the horizons.


\section{Discussion} \label{Con}
%
In this paper we have considered the constrained dynamics of field theories on Kerr backgrounds 
with one or more horizons and have argued that the constraints of the theory will receive 
additional contributions from these horizons. We explicitly considered the example of the Maxwell field, 
for which the Gauss law constraint was shown to involve contributions from the horizon(s). 
Such surface contributions will not arise on spacelike surfaces of the background,
but they appear on horizons in part due to our inability to observe past the horizon, as 
well as the fact that gauge fields can in principle take on arbitrary values at the horizon 
provided gauge invariant quantities constructed from them remain finite. More precisely, 
it is the non-vanishing of the gauge parameters and their derivatives at the horizons which leads 
to a Gauss law constraint with surface contributions.

Some consequences of the modified Gauss law constraint can be determined by gauge 
fixing the theory. In Sec.~(\ref{Max}) we considered 
two gauges -- the radiation gauge and the axial gauge. For the radiation gauge considered 
in Eq.~(\ref{gf.con}), we chose the covariant generalization of the gauge adopted in 
flat space. Unsurprisingly, the Dirac brackets for $a_a$ and $\pi^b$ in Eq.~(\ref{gf.cdir}) 
are the covariant generalizations of the flat space result, involving the Green function
of the spacetime Laplacian operator. As we noted in our treatment of this gauge however, 
it would be useful to include additional surface terms at the horizons. To see this,
let us now consider the following gauge fixing function
\begin{equation}
	\Omega_4 = \mathcal{D}^b\left(\beta^{-1} a_b\right) + \beta^{-1}n^b a_b \vert_{\cal{H}} \, ,
	\label{gf.con3}
\end{equation}
in place of the expression in Eq.~(\ref{gf.con}), with $\Omega_3$ as given. Unlike $\Omega_4$ 
in Eq.~(\ref{gf.con}), this gauge function involves additional terms at the horizons. The time derivative of $\Omega_4$ gives
\begin{equation}
\dot{\Omega}_4 = \mathcal{D}_a \pi^a + n^a\pi^a\Big\vert_{\mathcal{H}} = \Omega_2 \approx 0
\end{equation}
Thus this constraint is a consistent choice.
Proceeding as before, we now find that the bracket of $\left[\Omega_2(x), \Omega_4(y)\right]$ is given by
\begin{equation}
\left[\Omega_2 (x), \Omega_4 (y) \right]_P =  
	 \mathcal{D}^x_{a}\left(\beta^{-1}(x) \mathcal{D}_x^{a}\delta(x,y)\right) 
	 + \beta^{-1}(x) n_a^x \mathcal{D}_a^x \delta(x,y) \vert_{\cal{H}} \, .
\label{gf.con4}
\end{equation}
From the Schwarz inequality, it follows that the surface term in Eq.~(\ref{gf.con4}) does not vanish.
The resulting Dirac bracket will require the Green function for the operator involved in Eq.~(\ref{gf.con4}), which has a non-trivial surface contribution. This Green function and its derivatives do not vanish at the horizons. Hence the horizons will affect the 
Dirac brackets and the dynamics of the theory. 

A related point concerns the expression for the scalar potential $\phi$ following the axial gauge of Eq.~(\ref{gf2.gf}) 
\begin{align}
 & \mathcal{D}_b(f^{-2}\omega^b \omega_a \pi^a + \nu^b \nu_a \pi^a) 
 + \left[ n_b \nu^b \nu_a \pi^a\right]\vert_{\cal{H}}\notag\\
& \quad = \mathcal{D}_b\left(\beta^{-1} \mu^b \mu^a \left(\mathcal{D}^a \phi 
+ \alpha f_{ac}\omega^c\right) \right) + \beta^{-1} n_b \mu^b \mu^a \left(\mathcal{D}_a \phi 
+  \alpha f_{ac}\omega^c\right)\vert_{\cal{H}}\,,
\label{gf.axph}
\end{align}   
where we made use of Eq.~(\ref{gf2.proj2}). From Eq.~(\ref{gf.axph}) 
it also follows that $\phi$ depends non-trivially on $\pi^b$ at the horizon. 
Thus in general, the horizon correction in the Gauss law constraint will manifest 
in the dependent variables of the theory following gauge fixing.

Another implication of the Gauss law constraint involves the charges and currents 
on Kerr spacetimes. We noted in Sec.~\ref{CC} that the horizon correction in the 
Gauss law can be identified with the induced surface charge on the horizon of a 
black hole. This term was considered previously in the literature through boundary 
conditions on the normal component of the electric field. In addition, Maxwell's 
equations further imply an induced surface current as a consequence of the induced 
surface charge, which is related to components of the magnetic field parallel to 
the horizon. Thus corrections to the Gauss law constraint resulting from Killing 
horizons of the background lead to a natural identification of an induced surface 
charge and induced surface current in Eq.~(\ref{el.prp}) and Eq.~(\ref{mag.pll}) 
respectively. The induced surface charge in particular implies the vanishing of 
electric flux lines on the horizon, which was also noted in Sec.~\ref{CC}. 

Finally, we note that the BRST formalism provides an interesting and powerful means 
to investigate quantized fields in the Hamiltonian framework. Following our analysis in this paper 
and in~\cite{Fernandes:2016imn}, it can be argued that the BRST charge operator will involve the 
additional surface terms contained in the constraints. Thus the physical states defined by the 
cohomology of the BRST charge will have 
to satisfy non-trivial conditions on the horizon. We leave the investigation of this and related
questions for future work.

\appendix

\section{Derivation of the Dirac bracket $\left[\phi(x),\pi^b(y) \right]_{D}$} \label{app.der}
The second Dirac bracket provided in Eq.~(\ref{gf2.dir}) is given by  
\begin{align}
\left[\phi(x),\pi^b(y) \right]_{D} & = \mathcal{D}_a^y \left( \alpha(y) q(x,y) (\mu^a(y) \omega^b(y)  - \mu^b(y) \omega^a(y))\right)\notag\\
&\qquad \qquad + n_a^y \alpha(y) q(x,y) (\mu^a(y) \omega^b(y) )\vert_{\cal{H}} \,  \label{gf2.db2}
\end{align}
Here we provide its derivation to elaborate on the surface term. From Eq.~(\ref{gf.dir}) we have
\begin{equation}
	\left[\phi(x),\pi^b(y) \right]_{D} = - \int \limits_{\Sigma} dV_u \int \limits_{\Sigma} dV_v 
	\left[\phi(x), \,\Omega_\alpha(u)\right]_P C^{-1}_{\alpha\beta}\left(u, v\right) \left[\Omega_\beta(v)\,, \pi^b(y)\right]_P\,.
	\label{gf.dir2}
\end{equation} 
where we made use of the fact that $ \left[\phi(x),\pi^b(y) \right]_{P} = 0$. The expression in Eq.~(\ref{gf.dir2}) simplifies to
\begin{equation}
	\left[\phi(x),\pi^b(y) \right]_{D} = - \int \limits_{\Sigma} dV_u \int \limits_{\Sigma} dV_v 
	\left[\phi(x), \,\Omega_1(u)\right]_P C^{-1}_{14}\left(u, v\right) \left[\Omega_4(v)\,, \pi^b(y)\right]_P\,.
	\label{gf.dir22}
\end{equation} 
Eqs.~(\ref{gf2.gf}) and (\ref{gf2.inv}) can now be used to find the expression
\begin{align}
\left[\phi(x),\pi^b(y) \right]_{D} &= - \int \limits_{\Sigma} dV_u \delta(x,u) q(u,v) 
\,  \int \limits_{\Sigma} dV_v \alpha(v)(\mu^a(v)\omega^b(v) - \mu^b(v)\omega^a(v)) \mathcal{D}_a^v \delta(v,y) \notag\\
&= -\int \limits_{\Sigma} dV_v \, q(x,v)  \alpha(v)(\mu^a(v)\omega^b(v) - \mu^b(v)\omega^a(v)) \mathcal{D}_a^v \delta(v,y) \notag\\
&= \oint \limits_{\partial \Sigma} da_v \, \delta(v,y) n_a^v q(x,v) \alpha(v)(\mu^a(v)\omega^b(v)) \notag\\
& \quad \quad + \int \limits_{\Sigma} dV_v \delta(v,y) \mathcal{D}_a^v\left(\, q(x,v)  \alpha(v)(\mu^a(v)\omega^b(v) - \mu^b(v)\omega^a(v))\right) \, .
	\label{gf.db2r}
\end{align}
We made use of $n_a \omega^a = 0$ in the last equality of Eq.~(\ref{gf.db2r}).
Recalling that the brackets are in fact densities which need to be integrated over the hypersurface for both 
$x$ and $y$, we can express the result of Eq.~(\ref{gf.db2r}) as Eq.~(\ref{gf2.db2}).

\section{Axial gauge functions in Boyer-Lindquist coordinates} \label{A}
We will now explicitly derive the function $q(x,y)$ which appears in Eq.~(\ref{gf.q}). 
The Maxwell field is assumed to be defined on the Kerr background, for which we will 
adopt the usual Boyer-Lindquist coordinates $(t\,,r\,,\theta\,,\phi)$
\begin{align}
ds^2_{BL} = -\left(\frac{\Delta - a^2 \text{sin}^2\theta}{\rho^2}\right) dt^2 &+ \frac{2 \, a\, \text{sin}^2\theta}{\rho^2} \left(\Delta - r^2 - a^2\right) dt d\phi + \frac{\text{sin}^2\theta}{\rho^2} \left((r^2 + a^2)^2 - \Delta a^2 \text{sin}^2\theta\right) d\phi^2  \notag\\
& \qquad + \frac{\rho^2}{\Delta}dr^2 + \rho^2 d \theta^2 \, ,
\label{BL.met}
\end{align}
where 
\begin{equation}
\Delta = r^2 - 2 M r + a^2 \,, \qquad \rho^2 = r^2 + a^2 \text{cos}^2\theta \, ,
\label{BL.Dr}
\end{equation}
with $M$ being the mass of the black hole and $a$ the angular momentum per unit mass. In these coordinates the $r-\theta$ plane comprise the integral 2-submanifold orthogonal to both $t$ and $\phi$. In Sec.~\ref{Max} we noted that the axial gauge analysis could be carried out for any of the basis vectors of this submanifold. Here we will demonstrate this by deriving the function $q(x,y)$ separately for the cases 
$\mu^a = (\partial_r)^a$
and $\mu^a = (\partial_{\theta})^a$. From the inverse metric in Boyer-Lindquist 
coordinates, we have
\begin{equation}
(\partial_r)^a  = \left(0,\frac{\sqrt{\Delta}}{\rho},0,0\right) \,, \qquad (\partial_{\theta})^a = \left(0,0,\rho^{-1},0\right) \, .
\label{BL.base}
\end{equation}
Likewise, the metric components of Eq.~(\ref{BL.met}) provide the following definitions
\begin{align}
\lambda^2 &= - \frac{\Delta - a^2 \text{sin}^2\theta}{\rho^2} \notag\\
\alpha f^2 &= -\frac{a\, \text{sin}^2\theta}{\rho^2} \left(\Delta - r^2 - a^2\right) \notag\\
f^2 &=  \frac{\text{sin}^2\theta}{\rho^2} \left((r^2 + a^2)^2 - \Delta a^2 \text{sin}^2\theta\right) \, ,
\label{BL.sl}
\end{align}
as well as the following expressions for $\beta$ and $\sqrt{h}$ 
\begin{align}
\beta &= \sqrt{- (\lambda^2 + \alpha^2 f^2)} = \left( 1+ \frac{4 M r \left(a^2+r^2\right)}{\Delta \left(a^2 \cos (2 \theta)+a^2+2 r^2\right)}\right)^{-\frac{1}{2}} \notag\\
\sqrt{h} &= \frac{f \rho^2}{\sqrt{\Delta}} 
\label{BL.fol}
\end{align}
Since Eq.~(\ref{gf.q}) involves a delta function source, it will be convenient to first re-express it in terms of a second-order differential equation. For the case where $\mu^a = (\partial_r)^a$, Eq.~(\ref{gf.q}) can be explicitly rewritten as
\begin{equation}
\frac{\sqrt{\Delta(r',\theta')}}{\rho(r',\theta')} \partial_{r'} (\partial_{r'} l(\vec{r},\vec{r}')) = -\frac{1}{\sqrt{h(r',\theta')}}\delta(r -r')\delta(\theta -\theta')\delta(\phi -\phi')\, ,
\label{app.qr}
\end{equation}
where we have chosen $q(\vec{r},\vec{r}') = \partial_{r'} l(\vec{r},\vec{r}')$ and have considered the source at a fixed point $\vec{r}$. We now assume the following ansatz 
\begin{equation}
l(\vec{r},\vec{r}') = l(r,\theta, r')\delta(\theta -\theta')\delta(\phi -\phi') \, ,
\end{equation}
which simplifies Eq.~(\ref{app.qr}) to
\begin{equation}
\delta(\theta -\theta')\frac{\sqrt{\Delta(r',\theta')}}{\rho(r',\theta')} \partial_{r'} (\partial_{r'} l(r,\theta, r')) = -\frac{1}{\sqrt{h(r',\theta')}}\delta(r -r') \delta(\theta -\theta')\, ,
\label{app.qr2}
\end{equation}
The solution for $l(r,\theta, r')$ follows by first considering the homogeneous equation $\partial_{r'} (\partial_{r'} R(r,\theta, r')) = 0$, whose general solution is
\begin{equation}
R(r,\theta, r') = C_1(r,\theta) + C_2(r,\theta) r'
\end{equation}
Denoting the horizon radius as $r_H$, the solution $C_2(r,\theta) r'$ is valid only in the region $r_H \le r' < r$ (since it diverges in the region $r'>r$). The solution $C_1(r)$ is valid everywhere on $\Sigma$. By matching these solutions at the point $r=r'$, the general solution for $l(r,r')$ can then be written as
\begin{align}
l(r,\theta , r') &= C(r,\theta) r'   \qquad (r' < r) \notag\\
        &= C(r,\theta) r     \,  \qquad (r' > r) \label{BL.gs}
\end{align} 
Substituting this solution in Eq.~(\ref{app.qr2}), integrating $\theta'$ over its entire range and $r'$ from $r-\epsilon$ to $r+\epsilon$, we find that the constant $C(r, \theta)$ is given by
\begin{equation}
C(r,\theta) = \left(f(r,\theta) \rho(r,\theta)\right)^{-1} \, ,
\end{equation}
where $f$ and $\rho$ are defined in Eq.~(\ref{BL.sl}) and Eq.~(\ref{BL.Dr}) respectively. With Eq.~(\ref{BL.gs}), we now have the following general solution
\begin{align}
l(\vec{r},\vec{r}') &= \delta(\theta -\theta')\delta(\phi -\phi')\left(f(r,\theta) \rho(r,\theta)\right)^{-1} r'   \qquad (r' < r) \notag\\
        &= \delta(\theta -\theta')\delta(\phi -\phi')\left(f(r,\theta) \rho(r,\theta)\right)^{-1} r     \,  \qquad (r' > r) \, .
\end{align}
Differentiating this solution with respect to $r'$ gives
\begin{equation}
q(\vec{r},\vec{r}') = \frac{1}{f(r,\theta) \rho(r,\theta)} \Theta \left(r - r' \right) \delta(\theta -\theta')\delta(\phi -\phi') \, ,
\label{sol.qr}
\end{equation}
where $\Theta \left(r - r' \right)$ has the property that it is $1$ when $r_H < r' <r$ and $0$ elsewhere.\\
We can also consider the case where $\mu^a = (\partial_{\theta})^a$ in Eq.~(\ref{gf.q}). In this case, Eq.~(\ref{app.qr}) becomes
\begin{equation}
\frac{1}{\rho(r',\theta')} \partial_{\theta'} (\partial_{\theta'} l(\vec{r},\vec{r}')) = -\frac{1}{\sqrt{h(r',\theta')}}\delta(r -r')\delta(\theta -\theta')\delta(\phi -\phi')\, ,
\label{app.qt}
\end{equation}
By using the ansatz
 \begin{equation}
l(\vec{r},\vec{r}') = l(r,\theta, \theta')\delta(r -r')\delta(\phi -\phi') \, 
\end{equation}
and performing the analogous procedure described above, we find the following general solution for $q(\vec{r},\vec{r}')$
\begin{equation}
q(\vec{r},\vec{r}') = \frac{\sqrt{\Delta(r,\theta)}}{f(r,\theta) \rho(r,\theta)} \Theta \left(\theta - \theta' \right) \delta(r -r')\delta(\phi -\phi') \, ,
\label{sol.qt}
\end{equation}
$\Theta \left(\theta - \theta' \right)$ is now just the ordinary Heaviside step function.
Using the solutions given in Eq.~(\ref{sol.qr}) and Eq.~(\ref{sol.qt}), we can proceed to solve Eq.~(\ref{gf.p}) when $\mu^a$ is either $(\partial_r)^a$ or $(\partial_{\theta})^a$. We can alternatively rewrite Eq.~(\ref{gf.p}) as
\begin{equation}
\mu^a(y) \mathcal{D}^y_a\left(\beta^{-1}(y) \mu^a(y) \mathcal{D}_a^y p(x,y)\right) = \delta(x,y)
\end{equation}
and solve $p(x,y)$ using the procedure given above. The solutions for $p(x,y)$ about the Kerr background, when $\mu^a$ is either $(\partial_r)^a$ or $(\partial_{\theta})^a$, are not as simple as those of $q(x,y)$ and involve elliptic integrals. In the case of $\mu^a = (\partial_{\theta})^a$, the equation is  
\begin{equation}
\frac{1}{\rho(r',\theta')} \partial_{\theta'}\left(\beta^{-1}(r',\theta') \frac{1}{\rho(r',\theta')} \partial_{\theta'} p(\vec{r},\vec{r}')\right) = \frac{1}{\sqrt{h(r',\theta')}} \delta(r -r') \delta(\theta -\theta')\delta(\phi -\phi') \, ,
\end{equation}
whose solution is given by
\begin{equation}
 p(\vec{r},\vec{r}') = - \frac{\Delta(r,\theta)}{f(r,\theta) \rho(r,\theta)} \frac{F\left(\theta' \Big \vert \frac{a^2 \Delta}{(a^2 + r^2)^2}\right)}{2 (a^2 + r^2)} \delta(r -r')\delta(\phi -\phi') \, ,
\label{sol.pt}
\end{equation}
where $F\left(\theta' \big\vert k^2 \right)$ is the elliptic integral of the first kind. The solution of Eq.~(\ref{gf.p}) when $\mu^a = (\partial_r)^a$ involves elliptic integrals with much more complicated arguments and we were unable to find a simple expression as in Eq.~(\ref{sol.pt}). However in all cases, the functions $p$ and $q$ are curved spacetime generalizations of the $p(x,y)$ 
and $q(x,y)$ known in flat spacetime without boundaries~\cite{Hanson:1976cn}.

\section{Maxwell equations}\label{App.C}

Since the Hamiltonian $H$ involves an integral over the hypersurface $\Sigma$ which is 
orthogonal to $\chi\,,$ it is not 
obvious that it does generate time evolutions along $\xi$, as indicated in~\ref{Max}. As 
a check, we will here demonstrate that the Hamiltonian generates time evolution as specified, 
but is also consistent with Maxwell's equations when projected. By projecting 
$\nabla_a F^{ab}=0$ with Eq.~(\ref{gen.met}), one can find the following projected Maxwell equations 
\begin{align}
\pounds_{\chi}e_b&= - D_a(\beta f^{ab})\label{L.el} \\ 
\pounds_{\chi}f_{ab}&=-2D_{[a}\beta e_{b]}\,. \label{L.mag}
\end{align}
Turning our attention now to the Hamiltonian of Eq.~(\ref{U.fHam}), we find the following expressions upon evaluating the Poisson brackets
\begin{align}
\dot{\pi}^b = \left[\pi^b , H_T \right] &= \mathcal{D}_a(\beta f^{ab}) + \mathcal{D}_a\left(\alpha (\pi^a \omega^b - \pi^b \omega^a)\right) + \alpha n_a \pi^a \omega^b \Big \vert_{\cal{H}} \, , \label{H.el}\\
\dot{f}_{ab} = \left[f_{ab}, H_T \right] &= 2 \mathcal{D}_{[a} \beta \pi_{b]} + 2 \mathcal{D}_{[a}  \left( \alpha f_{b]c} \omega^c \right) \label{H.mag} \, .
\end{align}
It will be useful to note that since $\chi^a \omega_a = 0$, $\omega^c \nabla_c = \omega^c \mathcal{D}_c$ on any function or tensor.
Also from contracting Eq.~(\ref{chi.nLie}), we see that $\pounds_{\omega} \alpha = 0$. Thus, $\pounds_{\alpha \omega}$ of any spatially projected quantity can be written entirely in terms of the spatially projected covariant derivative. Let us first consider $\pounds_{\alpha \omega} f_{ab}$
\begin{align}
\pounds_{\alpha \omega} f_{ab} &= \alpha \omega^c \mathcal{D}_c f_{ab} + f_{ac} \mathcal{D}_b (\alpha \omega^c) + f_{cb} \mathcal{D}_a (\alpha \omega^c) \notag\\
&= 2 \alpha \omega^c \mathcal{D}_{[b}f_{a]c} + f_{ac} \mathcal{D}_b (\alpha \omega^c) + f_{cb} \mathcal{D}_a (\alpha \omega^c) \notag\\
&= -  2 \mathcal{D}_{[a}  \left( \alpha f_{b]c} \omega^c \right) \,, \label{lie.f}
\end{align}
where we made use of the Bianchi identity $\mathcal{D}_{[c}f_{ab]} = 0$ in going from the first equality to the second equality of Eq.~(\ref{lie.f}). Likewise, we find for $\pounds_{\alpha \omega} \pi^b$
\begin{align}
\pounds_{\alpha \omega}\pi^b &= \alpha \omega^c \mathcal{D}_c \pi^b - \pi^c\mathcal{D}_c\alpha \omega^b \notag\\
&= \mathcal{D}_c \left(\alpha (\omega^c \pi^b - \pi^c \omega^b)\right) - \alpha \omega^b \mathcal{D}_c\pi^c \, . \label{lie.p}
\end{align}
In going from the first equality to the final equation of Eq.~(\ref{lie.p}), we used the property that $\omega^c$ is Killing. 
Substituting Eq.~(\ref{lie.f}) in Eq.~(\ref{H.mag}) and Eq.~(\ref{lie.p}) in Eq.~(\ref{H.el}), we find
\begin{align}
\pounds_{\chi} \pi^{b} &= \mathcal{D}_a(\beta f^{ab}) +\alpha \omega^b \left(\mathcal{D}_a\pi^a + n_a \pi^a \Big \vert_{\cal{H}}\right) \approx \mathcal{D}_a(\beta f^{ab})  \, , \label{H.el2}\\
\pounds_{\chi} f_{ab} &= 2 \mathcal{D}_{[a} \beta \pi_{b]}  \label{H.mag2} \, .
\end{align}
Substituting Eq.~(\ref{H.mom}) in the above expressions, we get the projected Maxwell equations given in Eq.~(\ref{L.el}) and Eq.~(\ref{L.mag}). The derivation here should be contrasted with the analogous derivation on spherically symmetric backgrounds, where the time evolution vector is both Killing and orthogonal to the hypersurface. Thus, the foliation and time evolution as presented in this paper is consistent with the covariant Maxwell equations. 


\begin{thebibliography}{99}

\bibitem{Chandrasekhar:1985kt} 
  S.~Chandrasekhar,
  ``The mathematical theory of black holes,''
  OXFORD, UK: CLARENDON (1985) 646 P.


\bibitem{Hawking:1974sw} 
  S.~W.~Hawking,
  ``Particle Creation by Black Holes,''
  Commun.\ Math.\ Phys.\  {\bf 43}, 199 (1975)
  Erratum: [Commun.\ Math.\ Phys.\  {\bf 46}, 206 (1976)].
  doi:10.1007/BF02345020


\bibitem{Bekenstein:1973ur} 
  J.~D.~Bekenstein,
  ``Black holes and entropy,''
  Phys.\ Rev.\ D {\bf 7}, 2333 (1973).
  doi:10.1103/PhysRevD.7.2333


\bibitem{Balachandran:1994up} 
  A.~P.~Balachandran, L.~Chandar and A.~Momen,
  ``Edge states in gravity and black hole physics,''
  Nucl.\ Phys.\ B {\bf 461}, 581 (1996)
  doi:10.1016/0550-3213(95)00622-2


\bibitem{Balachandran:1995qa} 
  A.~P.~Balachandran, L.~Chandar and A.~Momen,
  ``Edge states in canonical gravity,''
  unpublished, gr-qc/9506006.


\bibitem{Carlip:1998wz} 
  S.~Carlip,
  ``Black hole entropy from conformal field theory in any dimension,''
  Phys.\ Rev.\ Lett.\  {\bf 82}, 2828 (1999)
  doi:10.1103/PhysRevLett.82.2828


\bibitem{Carlip:1999cy} 
  S.~Carlip,
  ``Entropy from conformal field theory at Killing horizons,''
  Class.\ Quant.\ Grav.\  {\bf 16}, 3327 (1999)
  doi:10.1088/0264-9381/16/10/322


\bibitem{Almheiri:2012rt} 
  A.~Almheiri, D.~Marolf, J.~Polchinski and J.~Sully,
  ``Black Holes: Complementarity or Firewalls?,''
  JHEP {\bf 1302}, 062 (2013)
  doi:10.1007/JHEP02(2013)062


\bibitem{Braunstein:2009my} 
  S.~L.~Braunstein, S.~Pirandola and K.~Życzkowski,
  ``Better Late than Never: Information Retrieval from Black Holes,''
  Phys.\ Rev.\ Lett.\  {\bf 110}, no. 10, 101301 (2013)
  doi:10.1103/PhysRevLett.110.101301


\bibitem{Balachandran:2013wsa} 
  A.~P.~Balachandran and S.~Vaidya,
  ``Spontaneous Lorentz Violation in Gauge Theories,''
  Eur.\ Phys.\ J.\ Plus {\bf 128}, 118 (2013)
  doi:10.1140/epjp/i2013-13118-9


\bibitem{Campiglia:2014yka} 
  M.~Campiglia and A.~Laddha,
  ``Asymptotic symmetries and subleading soft graviton theorem,''
  Phys.\ Rev.\ D {\bf 90}, no. 12, 124028 (2014)
  doi:10.1103/PhysRevD.90.124028


\bibitem{Strominger:2013lka} 
  A.~Strominger,
  ``Asymptotic Symmetries of Yang-Mills Theory,''
  JHEP {\bf 1407}, 151 (2014)
  doi:10.1007/JHEP07(2014)151


\bibitem{He:2014cra} 
  T.~He, P.~Mitra, A.~P.~Porfyriadis and A.~Strominger,
  ``New Symmetries of Massless QED,''
  JHEP {\bf 1410}, 112 (2014)
  doi:10.1007/JHEP10(2014)112


\bibitem{Strominger:2013jfa} 
  A.~Strominger,
  ``On BMS Invariance of Gravitational Scattering,''
  JHEP {\bf 1407}, 152 (2014)
  doi:10.1007/JHEP07(2014)152


\bibitem{He:2014laa} 
  T.~He, V.~Lysov, P.~Mitra and A.~Strominger,
  ``BMS supertranslations and Weinberg’s soft graviton theorem,''
  JHEP {\bf 1505}, 151 (2015)
  doi:10.1007/JHEP05(2015)151


\bibitem{Hawking:2016msc} 
  S.~W.~Hawking, M.~J.~Perry and A.~Strominger,
  ``Soft Hair on Black Holes,''
  Phys.\ Rev.\ Lett.\  {\bf 116}, no. 23, 231301 (2016)
  doi:10.1103/PhysRevLett.116.231301


\bibitem{Afshar:2016wfy} 
  H.~Afshar, S.~Detournay, D.~Grumiller, W.~Merbis, A.~Perez, D.~Tempo and R.~Troncoso,
  ``Soft Heisenberg hair on black holes in three dimensions,''
  Phys.\ Rev.\ D {\bf 93}, no. 10, 101503 (2016)
  doi:10.1103/PhysRevD.93.101503


\bibitem{Mirbabayi:2016axw} 
  M.~Mirbabayi and M.~Porrati,
  ``Dressed Hard States and Black Hole Soft Hair,''
  Phys.\ Rev.\ Lett.\  {\bf 117}, no. 21, 211301 (2016)
  doi:10.1103/PhysRevLett.117.211301


\bibitem{Hawking:2016sgy} 
  S.~W.~Hawking, M.~J.~Perry and A.~Strominger,
  ``Superrotation Charge and Supertranslation Hair on Black Holes,''
 JHEP {\bf 1705}, 161 (2017)
  doi:10.1007/JHEP05(2017)161


\bibitem{Tamburini:2017dig} 
  F.~Tamburini, M.~De Laurentis, I.~Licata and B.~Thidé,
  ``Twisted soft photon hair implants on Black Holes,''
 Entropy {\bf 19}, no. 9, 458 (2017)
  doi:10.3390/e19090458


\bibitem{Thorne:1986iy} 
  K.~S.~Thorne, R.~H.~Price and D.~A.~Macdonald,
  ``Black Holes: The Membrane Paradigm,''
  NEW HAVEN, USA: YALE UNIV. PR. (1986) 367p


\bibitem{Znajek:1978}
R.~L.~Znajek,
``The electric and magnetic conductivity of a Kerr hole''
Mon.\ Not.\ R.\ Astron.\ Soc.\ {\bf{185}} (4): 833-840 (1978)
doi:10.1093/mnras/185.4.833

\bibitem{Damour:1978cg} 
  T.~Damour,
  ``Black Hole Eddy Currents,''
  Phys.\ Rev.\ D {\bf 18}, 3598 (1978).
  doi:10.1103/PhysRevD.18.3598


\bibitem{Price:1986yy} 
  R.~H.~Price and K.~S.~Thorne,
  ``Membrane Viewpoint on Black Holes: Properties and Evolution of the Stretched Horizon,''
  Phys.\ Rev.\ D {\bf 33}, 915 (1986).
  doi:10.1103/PhysRevD.33.915


\bibitem{Blandford:1977ds} 
  R.~D.~Blandford and R.~L.~Znajek,
  ``Electromagnetic extractions of energy from Kerr black holes,''
  Mon.\ Not.\ Roy.\ Astron.\ Soc.\  {\bf 179}, 433 (1977).


\bibitem{Dirac:1950pj} 
  P.~A.~M.~Dirac,
  ``Generalized Hamiltonian dynamics,''
  Can.\ J.\ Math.\  {\bf 2}, 129 (1950).
  doi:10.4153/CJM-1950-012-1


\bibitem{Bergmann:1949zz} 
  P.~G.~Bergmann,
  ``Non-Linear Field Theories,''
  Phys.\ Rev.\  {\bf 75}, 680 (1949).
  doi:10.1103/PhysRev.75.680


\bibitem{Anderson:1951ta} 
  J.~L.~Anderson and P.~G.~Bergmann,
  ``Constraints in covariant field theories,''
  Phys.\ Rev.\  {\bf 83}, 1018 (1951).
  doi:10.1103/PhysRev.83.1018


\bibitem{Dirac-lect-1964} P. A. M. Dirac, ``Lectures on Quantum Mechanics'', Yeshiva University, New York,
1964.

\bibitem{Hanson:1976cn} 
  A.~J.~Hanson, T.~Regge and C.~Teitelboim,
  RX-748, PRINT-75-0141 (IAS,PRINCETON).


\bibitem{Henneaux:1992ig} 
  M.~Henneaux and C.~Teitelboim,
  ``Quantization of gauge systems,''
  Princeton, USA: Univ. Pr. (1992) 520 p


\bibitem{Dirac:1951zz} 
  P.~A.~M.~Dirac,
  ``The Hamiltonian form of field dynamics,''
  Can.\ J.\ Math.\  {\bf 3}, 1 (1951).
  doi:10.4153/CJM-1951-001-2


\bibitem{Arnowitt:1962hi} 
  R.~L.~Arnowitt, S.~Deser and C.~W.~Misner,
  ``The Dynamics of general relativity,''
  Gen.\ Rel.\ Grav.\  {\bf 40}, 1997 (2008)
  doi:10.1007/s10714-008-0661-1


\bibitem{Isenberg:1977ja} 
  J.~A.~Isenberg and J.~M.~Nester,
  ``Extension of the York Field Decomposition to General Gravitationally Coupled Fields,''
  Annals Phys.\  {\bf 108}, 368 (1977).
  doi:10.1016/0003-4916(77)90017-3


\bibitem{MacDonald:1982zz} 
  D.~MacDonald and K.~S.~Thorne,
  ``Black-hole electrodynamics - an absolute-space/universal-time formulation,''
  Mon.\ Not.\ Roy.\ Astron.\ Soc.\  {\bf 198}, 345 (1982).


\bibitem{Sorkin:1979ja} 
  R.~Sorkin,
  ``The Quantum Electromagnetic Field In Multiply Connected Space,''
  J.\ Phys.\ A {\bf 12}, 403 (1979).
  doi:10.1088/0305-4470/12/3/016


\bibitem{SheikhJabbari:1999xd} 
  M.~M.~Sheikh-Jabbari and A.~Shirzad,
  ``Boundary conditions as Dirac constraints,''
  Eur.\ Phys.\ J.\ C {\bf 19}, 383 (2001)
  doi:10.1007/s100520100590


\bibitem{Zabzine:2000ds} 
  M.~Zabzine,
  ``Hamiltonian systems with boundaries,''
  JHEP {\bf 0010}, 042 (2000)
  doi:10.1088/1126-6708/2000/10/042


\bibitem{Balachandran:1993tm} 
  A.~P.~Balachandran, L.~Chandar, E.~Ercolessi, T.~R.~Govindarajan and R.~Shankar,
  ``Maxwell-Chern-Simons electrodynamics on a disk,''
  Int.\ J.\ Mod.\ Phys.\ A {\bf 9}, 3417 (1994).
  doi:10.1142/S0217751X94001357


\bibitem{Balachandran:1992qg} 
  A.~P.~Balachandran and P.~Teotonio-Sobrinho,
  ``The Edge states of the BF system and the London equations,''
  Int.\ J.\ Mod.\ Phys.\ A {\bf 8}, 723 (1993)
  doi:10.1142/S0217751X9300028X


\bibitem{Fernandes:2016imn} 
  K.~Fernandes, S.~Ghosh and A.~Lahiri,
  ``Constrained field theories on spherically symmetric spacetimes with horizons,''
  Phys.\ Rev.\ D {\bf 95}, no. 4, 045012 (2017)
  doi:10.1103/PhysRevD.95.045012

\bibitem{Bhattacharya:2011dq} 
  S.~Bhattacharya and A.~Lahiri,
  ``No hair theorems for stationary axisymmetric black holes,''
  Phys.\ Rev.\ D {\bf 83}, 124017 (2011)
  doi:10.1103/PhysRevD.83.124017
  



\bibitem{Bojowald:2010qpa} 
  M.~Bojowald,
``Canonical Gravity and Applications: Cosmology, Black Holes, and Quantum Gravity,''
Cambridge University Press, New York, USA (2010) 312 p
  
\bibitem{Benguria:1976in} 
R.~Benguria, P.~Cordero and C.~Teitelboim,
``Aspects of the Hamiltonian Dynamics of Interacting Gravitational Gauge 
and Higgs Fields with Applications to Spherical Symmetry,''
Nucl.\ Phys.\ B {\bf 122}, 61 (1977).
doi:10.1016/0550-3213(77)90426-6



\bibitem{Ottewill:2012aj} 
  A.~C.~Ottewill and P.~Taylor,
  ``Static Kerr Green's Function in Closed Form and an Analytic Derivation of the 
  Self-Force for a Static Scalar Charge in Kerr Space-Time,''
  Phys.\ Rev.\ D {\bf 86}, 024036 (2012)
  doi:10.1103/PhysRevD.86.024036

\bibitem{Fernandes:2016sue} 
  K.~Fernandes and A.~Lahiri,
  Class.\ Quant.\ Grav.\  {\bf 34}, no. 17, 175004 (2017)
  doi:10.1088/1361-6382/aa7f61
  [arXiv:1601.01442 [gr-qc]].


\bibitem{Hanni:1973fn} 
  R.~S.~Hanni and R.~Ruffini,
  Phys.\ Rev.\ D {\bf 8}, 3259 (1973).
  doi:10.1103/PhysRevD.8.3259


\bibitem{Parikh:1997ma} 
  M.~Parikh and F.~Wilczek,
  Phys.\ Rev.\ D {\bf 58}, 064011 (1998)
  doi:10.1103/PhysRevD.58.064011
  [gr-qc/9712077].


\bibitem{Granot:2015xba} 
  J.~Granot, T.~Piran, O.~Bromberg, J.~L.~Racusin and F.~Daigne,
  Space Sci.\ Rev.\  {\bf 191}, no. 1-4, 471 (2015)
  doi:10.1007/s11214-015-0191-6
  [arXiv:1507.08671 [astro-ph.HE]].


\bibitem{Piran:2005qu} 
  T.~Piran,
  AIP Conf.\ Proc.\  {\bf 784}, 164 (2005)
  doi:10.1063/1.2077181
  [astro-ph/0503060].
  


%
%
%
\end{thebibliography}
\end{document}